\documentclass[12pt,a4]{article}
\newcommand{\text}{\rm}
\begin{document}
%%%%%%%%%%%%%%%%%%%%%%%%%%%%%%%%%%%%%%%%%%%
%\documentclass[12pt,a4]{article}
%%%%%%%%%%%%%%%%%%%%%%%%%%%%
%\usepackage{sw20lart}
%
%
%\input tcilatex
%\QQQ{Language}{
%American English
%}
%
%\begin{document}
%%%%%%%%%%%%%%%%%%%%%%%%%%%%%%%%%%%%%%%%%%%%%%%%%%
%%%%%%%%%%%%%%%%%%%%%%%%%%%%%%%%%%%%%%%%%%%%%%%%%%

\title{\textbf{Large-mass behaviour of loop variables in abelian
Maxwell--Chern--Simons theory }}
\author{\textbf{V.E.R. Lemes, C.A. Linhares, S.P. Sorella, L.C.Q.Vilar} \and %
\vspace{2mm} \\
%EndAName
Instituto de F\'{\i}sica,\\
Universidade do Estado do Rio de Janeiro\\
Rua S\~{a}o Francisco Xavier, 524\\
20550-013 Rio de Janeiro, Brazil \\
\vspace{2mm} \and \textbf{UERJ/DFT-02/98}\vspace{2mm}\newline \and \textbf{%
PACS: 11.10.Gh}}
\maketitle

\begin{abstract}
The large-mass behaviour of loop variables in Maxwell--Chern--Simons theory
is analysed by means of a gauge-field transformation which allows to reset
the Maxwell--Chern--Simons action to pure Chern--Simons.

\setcounter{page}{0}\thispagestyle{empty}
\end{abstract}

\vfill\newpage\ \makeatother
\renewcommand{\theequation}{\thesection.\arabic{equation}}

\section{Introduction\-}

In recent works \cite{lett1,lett2} it has been established that
three-dimensional gauge theories in the presence of the topological
Chern--Simons term can be cast in the form of a pure Chern--Simons action
through a local covariant redefinition of the gauge connection. For
instance, in the case of the standard Yang--Mills term $\int FF$, we get 
\begin{equation}
\mathcal{S}_{\text{CS}}(A)+\frac 1{4m}\text{tr}\int d^3xF_{\mu \nu }F^{\mu
\nu }=\mathcal{S}_{\text{CS}}(\widehat{A})\;,  \label{reab}
\end{equation}
with 
\begin{equation}
\widehat{A}_\mu =A_\mu +\sum_{n=1}^\infty \frac 1{m^n}\vartheta _\mu ^n\;,
\label{nl}
\end{equation}
and 
\begin{equation}
\mathcal{S}_{\text{CS}}(A)=\frac 12\text{tr}\int d^3x\varepsilon ^{\mu \nu
\rho }\left( A_\mu \partial _\nu A_\rho +\frac 23gA_\mu A_\nu A_\rho \right)
\;.  \label{cs}
\end{equation}
The two parameters $g,m$ in the expressions (\ref{reab}), (\ref{cs}) can be
identified respectively with the gauge coupling and with the topological
mass. The coefficients $\vartheta _\mu ^n$ in eq.(\ref{nl}) have been proven 
\cite{lett2} to transform covariantly under gauge transformations and can be
expressed in terms of the curvature $F_{\mu \nu }$ and its covariant
derivatives. This implies that the redefined field $\widehat{A}_\mu $ still
is a connection. This property has led to an attractive geometrical
interpretation of the Chern--Simons term as a kind of topological generator
for the classical Yang--Mills-type actions \cite{lett2}.

The existence of the transformation (\ref{nl}) has been exploited also at
the quantum level. Several results have been obtained for the quantum
effective actions of different systems. In the case of the massive
topological Yang--Mills (\ref{reab}), the redefinition (\ref{nl}) has
allowed for a purely algebraic proof of the ultraviolet finiteness of the
model \cite{lett2}. Moreover, in the case of the effective action
corresponding to the abelian fermionic determinant of a massive
two-component spinor field, eq.(\ref{nl}) has been extended at the quantum
level in order to account for the nonlocal quantum corrections \cite{lett3}.
As a consequence, it has been possible to prove that the infinite number of
one-loop diagrams corresponding to the perturbative expansion of the
fermionic determinant can be reabsorbed into the pure Chern--Simons, up to a
field redefinition \cite{lett3}.

On the other side, it is known that pure Chern--Simons theory can be
recovered as the infinite-mass limit $m\rightarrow \infty $ of the massive
topological Yang--Mills action (\ref{reab}). This property has been proven
to hold for both the 1PI effective action \cite{r1} and for the vacuum
expectation value of the Wilson loop \cite{rr}. We expect thus that, for a
finite large value of the mass parameter $m$ (\textit{i.e.}, a\textit{\ }%
low-energy\textit{\ }regime)\textit{,} the effects of the presence of the
Yang--Mills term show up in the form of a power series in $1/m$. 
Furthermore, being the field redefinition (\ref{nl}) a local expansion in $1/m$, it gives us a natural way to deal with the large-mass corrections
which affect the vacuum expectation value of the observables, \textit{i.e., }%
of the nontrivial gauge-invariant quantities.

This is the aim of this article. In particular, we shall investigate whether
the redefinition (\ref{nl}) can be used as an effective computational tool
in order to characterize the large-mass behaviour of the observables. In
this work we shall restrict ourselves to the abelian case. The main idea is
to use eq.(\ref{nl}) as a change of variables in the path integral. In doing
this, one picks up the Jacobian of the transformation and one has to
re-express the observable under consideration in terms of the redefined
field $\widehat{A}_\mu $, which, being now a power series in $1/m$, will
systematically produce a local expansion of the observable in powers of $1/m$%. However, such a change of variables in the path integral leads the
Boltzmann weight to take the form of the pure Chern--Simons action. This
procedure may therefore have the practical advantage of performing
computations by making use of the Chern--Simons propagator 
\begin{equation}
\mathcal{G}_{\mu \nu }^{\text{CS}}=-\frac 1{4\pi }\varepsilon _{\mu \nu \rho
}\partial ^\rho \frac 1{\left| x-y\right| }\;=\frac 1{4\pi }\varepsilon
_{\mu \nu \rho }\frac{(x-y)^\rho }{\left| x-y\right| ^3}\;,  \label{cs-pr}
\end{equation}
instead of the more complicated one corresponding to the quadratic part of
the action (\ref{reab}), \textit{i.e.,} 
\begin{equation}
\mathcal{G}_{\mu \nu }^{\text{MCS}}=\frac 1{4\pi }\varepsilon _{\mu \nu \rho
}\frac{(x-y)^\rho }{\left| x-y\right| ^3}+\frac m{4\pi }\frac{e^{-m\left|
x-y\right| }}{\left| x-y\right| }\left( g_{\mu \nu }-\frac 1m\varepsilon
_{\mu \nu \rho }\frac{(x-y)^\rho }{\left| x-y\right| ^2}(1+m\left|
x-y\right| \right) \;.\;  \label{mcs-pr}
\end{equation}
In other words, in the case of a large value of $m$, the field redefinition (%
\ref{nl}) will allow us to shift the mass dependence from the Boltzmann
weight directly to the observable, so that the expectation value can be
obtained by making use of the pure Chern--Simons propagator. As an
illustrating example of this set-up we shall use the field redefinition (\ref
{nl}) in the case of the abelian Maxwell--Chern--Simons theory in flat
space-time 
\begin{equation}
\mathcal{S}_{\text{MCS}}(A)=\frac 12\int d^3x\varepsilon ^{\mu \nu \rho
}A_\mu \partial _\nu A_\rho +\frac 1{4m}\int d^3xF_{\mu \nu }F^{\mu \nu }\;,
\label{mcs-th}
\end{equation}
in order to compute the large-mass corrections to the loop factor \cite
{hr,br} 
\begin{equation}
\mathcal{W}(\gamma ,\gamma ^{\prime })=e^{-\oint_\gamma dx^\mu \oint_{\gamma
^{\prime }}dy^\nu \left\langle A_\mu (x)A_\nu (y)\right\rangle }\;,
\label{loop}
\end{equation}
where $\gamma ,\gamma ^{\prime }$ are two distinct (nonintersecting) smooth
closed oriented curves (which define a two-component link $L(\gamma ,\gamma
^{\prime })$ \cite{gu}). As discussed by \cite{br}, the relevance of the
factor $\mathcal{W}(\gamma ,\gamma ^{\prime })$ is due to the fact that it
carries the information concerning the statistics of the quantum
fluctuations of the $(2+1)$-dimensional abelian Higgs model, thanks to a
random-walk representation for the gauge-invariant Green's functions. In
particular, by making use of the redefinition (\ref{nl}), we shall be able
to prove that all the expected corrections in the inverse of the mass
parameter $m$ are actually absent, provided $\gamma ,\gamma ^{\prime }$ are
two disjoint curves, one of which may wind around the other. In other words,
the double line integral $\oint_\gamma dx^\mu \oint_{\gamma ^{\prime
}}dy^\nu \left\langle A_\mu (x)A_\nu (y)\right\rangle $ computed with the
Maxwell--Chern--Simons action yields, in the large-mass limit, the linking
number $\chi (\gamma ,\gamma ^{\prime })$.

It is worth remarking here that this set-up could have useful applications
for the three-dimensional bosonization of fermionic systems for large value
of the fermion mass. We recall in fact that the large-mass expansion of the
so-called three-dimensional fermionic determinant is one of the basic
ingredients of our present understanding of three-dimensional bosonization 
\cite{bos}.

The paper is organized as follows. In Sect.2 we discuss the main properties
of the field redefinition (\ref{nl}) in the case of the abelian
Maxwell--Chern--Simons theory. In Sect.3 the details of the computation of
the double line integral $\oint_\gamma dx^\mu \oint_{\gamma ^{\prime
}}dy^\nu \left\langle A_\mu (x)A_\nu (y)\right\rangle $ will be given. In
Sect.4 we discuss the extension of this result to a more general class of
abelian actions as well as to a generic link $L(\gamma _1,...,\gamma _n)$.
Sect.5 is devoted to the case in which $\gamma $ and $\gamma ^{\prime }$
identify the same curve, expression (\ref{loop}) becoming there the
expectation value of the abelian Wilson loop \cite{p2,hr, coste, rr}. The
example of a planar loop will be reported in detail.

Although the present work deals with the large mass behaviour of the loop
variables, a simple framework for the case of the small-mass corrections is
provided in Appendix A.

\section{The Maxwell-Chern-Simons action}

As already mentioned, the abelian Maxwell--Chern--Simons action can be reset
to a pure Chern--Simons term 
\begin{equation}
\mathcal{S}_{\text{MCS}}(A)=\frac 12\int d^3x\,\varepsilon ^{\mu \nu \rho
}A_\mu \partial _\nu A_\rho +\frac 1{4m}\int d^3x\,F_{\mu \nu }F^{\mu \nu
}\;=\mathcal{S}_{\text{CS}}(\widehat{A})\;,  \label{mcs-red}
\end{equation}
through a suitable gauge field redefinition of the kind 
\begin{equation}
\widehat{A}_\mu =A_\mu +\sum_{n=1}^\infty \frac 1{m^n}\vartheta _\mu ^n\;.
\label{ga-red}
\end{equation}
As proven in ref.\cite{lett2} by using BRST cohomological techniques, the
coefficients $\vartheta _\mu ^n$ turn out to depend only on the field
strength and its derivatives. In the present abelian case their computation
is rather straightforward. For instance, the first six coefficients are
found to be 
\begin{eqnarray}
\vartheta _\mu ^1 &=&\frac 14\varepsilon _{\mu \nu \rho }F^{\nu \rho
}\;,\;\;\;\;\;\;\;\;\;\;\;\;\;\;\;\;\;\;\vartheta _\mu ^2=-\frac 18\partial
^\nu F_{\mu \nu }\;,  \nonumber  \label{six} \\
\;\vartheta _\mu ^3 &=&-\frac 1{32}\varepsilon _{\mu \nu \rho }\partial
^2F^{\nu \rho }\;,\;\;\;\;\;\;\;\;\;\;\;\;\;\vartheta _\mu ^4=\frac
5{128}\partial ^2\partial ^\nu F_{\mu \nu }\;,  \nonumber \\
\vartheta _\mu ^5 &=&\frac 7{512}\varepsilon _{\mu \nu \rho }\partial
^4F^{\nu \rho }\;,\;\;\;\;\;\;\;\;\;\;\;\;\;\vartheta _\mu ^6=-\frac{21}{1024%
}\partial ^4\partial ^\nu F_{\mu \nu }\;.\;  \label{six}
\end{eqnarray}
Although the higher-order coefficients can be easily obtained, the above
expressions give us a simple and clear understanding of the general
properties of the $\vartheta _\mu ^n$ 's. They can be summarized as follows:

\begin{itemize}
\item  the coefficients $\vartheta _\mu ^n$ are divergenceless, \textit{i.e.,%
} 
\begin{equation}
\partial ^\mu \vartheta _\mu ^n=0\;,  \label{div}
\end{equation}

\item  they are gauge invariant and depend linearly on the gauge field $%
A_\mu $.
\end{itemize}

As one can easily understand, these properties follow from the abelian
character of the Maxwell--Chern--Simons action (\ref{mcs-red}). They will
turn out to be of great relevance in order to compute the large mass
behaviour of $\mathcal{W}(\gamma ,\gamma ^{\prime })$. In particular, from
equation (\ref{div}) it follows that the general form of the field
transformation (\ref{ga-red}) can be written in terms of the two transverse
projectors $\varepsilon _{\mu \nu \rho }\partial ^\rho $ and $(g_{\mu \nu
}\partial ^2-\partial _\mu \partial _\nu )\;$as 
\begin{equation}
\widehat{A}_\mu =A_\mu +\frac 1m\varepsilon _{\mu \nu \rho }f(\partial
^2/m^2)\partial ^\nu A^\rho -\frac 1{m^2}h(\partial ^2/m^2)(g_{\mu \nu
}\partial ^2-\partial _\mu \partial _\nu )A^\nu \;,  \label{gen-red}
\end{equation}
where $f$ and $h$ are power series in the three-dimensional laplacian 
\begin{eqnarray}
f(\partial ^2/m^2) &=&\frac 12-\frac 1{16}\frac{\partial ^2}{m^2}+\frac
7{256}\frac{\partial ^4}{m^4}+\;.....\;,  \label{fh} \\
h(\partial ^2/m^2) &=&-\frac 18+\frac 5{128}\frac{\partial ^2}{m^2}-\frac{21%
}{1024}\frac{\partial ^4}{m^4}+\;....\;.  \nonumber
\end{eqnarray}
Observe that from eq.(\ref{div}) it follows that the two gauge connections $%
\widehat{A}_\mu $ and $A_\mu $ have the same divergence, 
\begin{equation}
\partial ^\mu \widehat{A}_\mu =\partial ^\mu A_\mu \;,  \label{s-div}
\end{equation}
which implies that, in a covariant Lorentz-type gauge condition, the
gauge-fixing term remains unchanged when one moves from $A_\mu \;$to $%
\widehat{A}_\mu $.

Let us give here, for later use, the coefficients of the inverse
transformation (\ref{six}) which has, of course, the same general form of
eq.(\ref{gen-red}): 
\begin{equation}
A_\mu =\widehat{A}_\mu +\frac 1m\varepsilon _{\mu \nu \rho }\widehat{f}%
(\partial ^2/m^2)\partial ^\nu \widehat{A}^\rho -\frac 1{m^2}\widehat{h}%
(\partial ^2/m^2)(g_{\mu \nu }\partial ^2-\partial _\mu \partial _\nu )%
\widehat{A}^\nu \;,  \label{i-gen-red}
\end{equation}
with 
\begin{eqnarray}
\widehat{f}(\partial ^2/m^2) &=&-\frac 12+\frac 5{16}\frac{\partial ^2}{m^2}-%
\frac{63}{256}\frac{\partial ^4}{m^4}+\;....\;,  \label{i-six} \\
\widehat{h}(\partial ^2/m^2) &=&\frac 38-\frac{35}{128}\frac{\partial ^2}{m^2%
}+\;\frac{231}{1024}\frac{\partial ^4}{m^4}....\;.  \nonumber
\end{eqnarray}
It should also be remarked that eqs.(\ref{gen-red}) and (\ref{i-gen-red}),
being linear in the fields $A_\mu $, $\hat{A}_\mu ,$ imply that the Jacobian 
$\det (\delta A_\nu /\delta \widehat{A}_\mu )$ corresponding to the change
of variables $A_\mu \rightarrow \widehat{A}_\mu \;$is field independent, and
therefore it does not contribute to the transformed measure $\mathcal{D}%
\widehat{A}\;$in the path integral.

We are now ready to evaluate the large mass effects to the double line
integral of the expression (\ref{loop})$.$ This will be the task of the next
section.

\section{Computation of the double line integral}

In order to compute the line integral $\oint_\gamma dx^\mu \oint_{\gamma
^{\prime }}dy^\nu \left\langle A_\mu (x)A_\nu (y)\right\rangle $ in the
Maxwell--Chern--Simons theory we have first to fix the gauge. Adopting a
transverse Landau gauge, we can write 
\begin{equation}
\oint_\gamma \oint_{\gamma ^{\prime }}\left\langle A(x)A(y)\right\rangle _{%
\text{MCS}}=\frac{\int \mathcal{D}A\mathcal{D}b\oint_\gamma dx^\mu
\oint_{\gamma ^{\prime }}dy^\nu A_\mu (x)A_\nu (y)\;e^{i\left( \mathcal{S}_{%
\text{MCS}}(A)+\int d^3x\,b\partial A\right) }}{\int \mathcal{D}A\mathcal{D}%
b\;e^{i\left( \mathcal{S}_{\text{MCS}}(A)+\int d^3x\,b\partial A\right) }}\;,
\label{exp1}
\end{equation}
where $b$ is the Lagrange multiplier implementing the gauge condition. Let
us perform now the change of variables (\ref{i-gen-red})$.$ Moreover,
recalling that the corresponding Jacobian is field independent and that,
from eq.(\ref{s-div}), the Landau gauge condition is left invariant, we get 
\begin{equation}
\oint_\gamma dx^\mu \oint_{\gamma ^{\prime }}dy^\nu \left\langle A_\mu
(x)A_\nu (y)\right\rangle _{\text{MCS}}=\oint_\gamma dx^\mu \oint_{\gamma
^{\prime }}dy^\nu \left\langle A_\mu (\widehat{A}(x))A_\nu (\widehat{A}%
(y))\right\rangle _{\text{CS\ }},  \label{id}
\end{equation}
where 
\[
\oint_\gamma dx^\mu \oint_{\gamma ^{\prime }}dy^\nu \left\langle A_\mu (%
\widehat{A}(x))A_\nu (\widehat{A}(y))\right\rangle _{\text{CS}}\qquad \qquad
\qquad \qquad \qquad \qquad 
\]
\begin{equation}
=\frac{\int \mathcal{D}\widehat{A}\mathcal{D}b\oint_\gamma dx^\mu
\oint_{\gamma ^{\prime }}dy^\nu (\widehat{A}_\mu (x))A_\nu (\widehat{A}%
(y))\;e^{i\left( \mathcal{S}_{\text{CS}}(\widehat{A})+\int d^3x\,b\partial 
\widehat{A}\right) }}{\int \mathcal{D}\widehat{A}\mathcal{D}b\;e^{i\left( 
\mathcal{S}_{\text{CS}}(\widehat{A})+\int d^3x\,b\partial \widehat{A}\right)
}}.
\end{equation}
We see therefore that the expectation value of $\oint_\gamma \oint_{\gamma
^{\prime }}\left\langle A(x)A(y)\right\rangle _{\text{MCS}}$ in the
Maxwell--Chern--Simons theory can be obtained by computing the expectation
value of the transformed quantity $\oint_\gamma \oint_{\gamma ^{\prime
}}\left\langle A(\widehat{A}(x))A(\widehat{A}(y))\right\rangle _{\text{CS}}$
in the (topological) pure Chern--Simons theory. Therefore 
\[
\oint_\gamma dx^\mu \oint_{\gamma ^{\prime }}dy^\nu \left\langle A_\mu (%
\widehat{A}(x))A_\nu (\widehat{A}(y))\right\rangle _{\text{CS}%
}\;\;\;\;\;\;\;\;\;\;\;\;\;\;\;\;\;\;\;\;\;\;\;\;\;\;\;\;\;\;\;\;\;\;\;\;\;%
\;\;\;\;\;\;\; 
\]
\begin{equation}
\;\;\;\;\;\;\;\;\;\;\;\;\;=\oint_\gamma dx^\mu \oint_{\gamma ^{\prime
}}dy^\nu \Omega _{\mu \sigma }(x)\Omega _{\nu \lambda }(y)\left\langle 
\widehat{A}^\sigma (x)\widehat{A}^\lambda (y)\right\rangle _{\text{CS}}\;,
\label{con}
\end{equation}
with $\Omega _{\mu \sigma }(x)$ given by 
\begin{equation}
\Omega _{\mu \sigma }(x)=\left( g_{\mu \sigma }+\frac 1m\varepsilon _{\mu
\rho \sigma }\widehat{f}(\partial _x^2/m^2)\partial _x^\rho -\frac 1{m^2}%
\widehat{h}(\partial _x^2/m^2)(g_{\mu \sigma }\partial ^2-\partial _\mu
\partial _\sigma )_x\right) \;.  \label{om}
\end{equation}
In order to evaluate expression (\ref{con}) let us recall that the Landau
propagator of the pure Chern--Simons theory, 
\begin{equation}
\left\langle \widehat{A}^\sigma (x)\widehat{A}^\lambda (y)\right\rangle _{%
\text{CS}}=-\frac 1{4\pi }\varepsilon ^{\sigma \lambda \tau }\partial _\tau
\frac 1{\left| x-y\right| }\;,  \label{cs-p-d}
\end{equation}
is transverse, 
\begin{equation}
\partial _\sigma \left\langle \widehat{A}^\sigma (x)\widehat{A}^\lambda
(y)\right\rangle _{\text{CS}}=0\;,  \label{tr}
\end{equation}
and that, from 
\begin{equation}
\partial ^2\frac 1{\left| x-y\right| }=-4\pi \delta ^3(x-y)\;,\;  \label{lap}
\end{equation}
we get 
\begin{equation}
\partial ^2\left\langle \widehat{A}^\sigma (x)\widehat{A}^\lambda
(y)\right\rangle _{\text{CS}}=0\;,\;\;\;\;\;\;\;\text{for}\;\;\;x\neq y\;.
\label{imp-id}
\end{equation}
This last identity can be directly applied to eq.(\ref{con}), as $\gamma $
and $\gamma ^{\prime }$ are two disjoint (nonintersecting) curves. Therefore
the points $x$ and $y$ will never coincide. As a consequence, all the
laplacians in the factors $\Omega $ of eq.(\ref{con}) can be eliminated. The
same occurs for the terms containing the derivatives $\partial _\mu \partial
_\sigma $ and $\partial _\nu \partial _\lambda ,$ due to the transversality
of the Chern--Simons propagator. Expression (\ref{con}) thus reduces to 
\[
\oint_\gamma dx^\mu \oint_{\gamma ^{\prime }}dy^\nu \left\langle A_\mu (%
\widehat{A}(x))A_\nu (\widehat{A}(y))\right\rangle _{\text{CS}}\qquad \qquad
\qquad \qquad \qquad \qquad \qquad 
\]
\begin{eqnarray}
&=&-\frac 1{4\pi }\oint_\gamma dx^\mu \oint_{\gamma ^{\prime }}dy^\nu
\varepsilon _{\mu \nu \tau }\partial ^\tau \frac 1{\left| x-y\right| }\; 
\nonumber \\
&\;&-\frac 1{4\pi m}\oint_\gamma dx^\mu \oint_{\gamma ^{\prime }}dy^\nu
\varepsilon _{\mu \rho \sigma }\varepsilon _{\;\;\;\;\nu }^{\sigma \lambda
}\partial ^\rho \partial _\lambda \frac 1{\left| x-y\right| }\;  \nonumber \\
&&+\frac 1{16\pi m^2}\oint_\gamma dx^\mu \oint_{\gamma ^{\prime }}dy^\nu
\varepsilon _{\mu \rho \sigma }\varepsilon _{\nu \tau \lambda }\varepsilon
^{\sigma \lambda \alpha }\partial ^\rho \partial ^\tau \partial _\alpha
\frac 1{\left| x-y\right| }  \nonumber \\
\;\;\;\;\; &=&-\frac 1{4\pi }\oint_\gamma dx^\mu \oint_{\gamma ^{\prime
}}dy^\nu \varepsilon _{\mu \nu \tau }\partial ^\tau \frac 1{\left|
x-y\right| }\;,\;{}  \label{exp3}
\end{eqnarray}
where all the $1/m$-dependent terms turn out to identically vanish or to
yield a total derivative upon contraction of the $\varepsilon _{\mu \nu \rho
}$ factors. For the final result we therefore get \cite{gu} 
\begin{equation}
\oint_\gamma dx^\mu \oint_{\gamma ^{\prime }}dy^\nu \left\langle A_\mu
(x)A_\nu (y)\right\rangle _{\text{MCS}}=\chi (\gamma ,\gamma ^{\prime })\;,
\label{f-r}
\end{equation}
where $\chi (\gamma ,\gamma ^{\prime })$ is the linking number of the two
curves $\gamma $ and $\gamma ^{\prime }$. We may see, then, that, as
announced, the factor $\mathcal{W}(\gamma ,\gamma ^{\prime })\;$is not
affected by large-mass corrections in $1/m.$ As one can easily understand,
this is due to the fact that the two curves do not intersect each other. It
should also be remarked that the use of the transformation (\ref{gen-red})
has allowed us to perform the computations straightforwardly by making use
of the properties (\ref{tr}), (\ref{imp-id}) of the pure Chern-Simons
propagator.

\section{Generalization}

Following the algebraic cohomological set-up of refs.\cite{lett1,lett2}, it
follows that the above result (\ref{f-r}) can be easily extended to the case
in which we add to the Maxwell--Chern--Simons action (\ref{mcs-red}) higher
derivatives terms of the type 
\begin{equation}
\frac{\alpha _n}{m^{2n+1}}\int d^3xF_{\mu \nu }(\partial ^2)^nF^{\mu \nu
}\;\;,\;\;\;\;\;\;\;\frac{\beta _n}{m^{2n}}\int d^3x\varepsilon ^{\mu \nu
\rho }A_\mu \partial _\nu (\partial ^2)^nA_\rho \;,\;\;\;\;\;\;n\geq 1\;,
\label{gen}
\end{equation}
where $\alpha _n$, $\beta _n$ are arbitrary dimensionless parameters.

These terms, being quadratic in the gauge field $A_\mu $,$\;$can be
reabsorbed into the pure Chern--Simons action through a linear field
redefinition of the kind (\ref{gen-red})$.$ Everything works as before, with
the result that the double line integral $\oint_\gamma dx^\mu \oint_{\gamma
^{\prime }}dy^\nu \left\langle A_\mu (x)A_\nu (y)\right\rangle $ is not
affected, in the large-mass limit, by corrections in $1/m$, meaning that it
is in fact independent from the parameters $\alpha _n$, $\beta _n$. We
recall here that the terms of eq.(\ref{gen}), together with the
Maxwell--Chern--Simons action (\ref{mcs-th}), appear in the large-mass
expansion of the two-point Green's function (\textit{i.e., }of the spinor
vacuum polarization) of the effective action corresponding to the abelian
fermionic determinant of a two-component massive spinor \cite{spin}.

Finally, it is worth underlining that all the results established here can
be generalized straightforwardly to a generic line integral $I(\gamma
_1,...,\gamma _n)$ of the type 
\begin{equation}
I(\gamma _1,...,\gamma _n)=\oint_{\gamma _1}dx_1^{\mu _1}\oint_{\gamma
_2}dx_2^{\mu _2}\;.....\oint_{\gamma _n}dx_n^{\mu _n}\left\langle A_{\mu
_1}(x_1)......A_{\mu _n}(x_n)\right\rangle \;,  \label{i-int}
\end{equation}
where the curves $\gamma _{1,....,}\gamma _n$ belong to a $n$-component link 
$L(\gamma _1,...,\gamma _n)$.

\section{Large mass behaviour of the Wilson Loop}

In this Section we consider the degenerate case of the Wilson loop, which
amounts to computing, within the Maxwell--Chern--Simons context, the link
variable $I(\gamma ,\gamma ^{\prime })$ when $\gamma $ and $\gamma ^{\prime
} $ both refer to the same curve, that is, 
\begin{equation}
I(\gamma ,\gamma )=\oint_\gamma dx^\mu \oint_\gamma dy^\nu \left\langle
A_\mu (x)A_\nu (y)\right\rangle _{\text{MCS}}\;.  \label{wilsonl}
\end{equation}
It is worth reminding that the double line integral (\ref{wilsonl}) computed
in pure Chern--Simons is finite and can be defined as the writhing number of
the curve $\gamma $ \cite{gu,writhe}. Moreover, the authors of ref.\cite{rr}
have been able to show that (\ref{wilsonl}) computed in
Maxwell--Chern--Simons theory yields, in the infinite mass limit $%
m\rightarrow \infty $, the so-called self-linking number \cite
{selflink,gu,writhe}. They have also proven that, by means of a finite
renormalization, the self-linking can be converted into the writhing number,
thus recovering the previous infinite-mass limit results of \cite
{p2,hr,coste}.

However, up to our knowledge, the mass dependence of the abelian Wilson loop
for a finite large value of the mass parameter $m$ has not yet been
completely worked out. Our purpose here is to show how the present set-up
can be useful in evaluating the large mass contributions which affect the
expression (\ref{wilsonl}).

For large $m$, we perform once more the field redefinition (\ref{i-gen-red}%
), which allows us to use the Chern--Simons propagator. It leads to an
expansion of $I(\gamma ,\gamma )$ in $1/m$ similar to the one given in (\ref
{exp3}); however, we now cannot eliminate all $1/m$-dependent terms, because
in the present case the integration variables $x$ and $y$ both refer to
points along the same curve. The first few terms in the expansion are seen
to be 
\begin{eqnarray}
I(\gamma ,\gamma )\; &=&-\frac 1{4\pi }\oint_\gamma dx^\mu \oint_\gamma
dy^\nu \,\varepsilon _{\mu \nu \alpha }\partial _x^\alpha \frac 1{\left|
x-y\right| }  \nonumber  \label{wlm3} \\
&&-\frac 1m\oint_\gamma dx^\mu \oint_\gamma dy_\mu \;\delta ^3(x-y) 
\nonumber \\
&&-\frac 1{m^2}\oint_\gamma dx^\mu \oint_\gamma dy^\nu \,\varepsilon _{\mu
\nu \alpha }\partial _x^\alpha \delta ^3(x-y)  \label{wlm3} \\
&&+\frac 1{m^3}\oint_\gamma dx^\mu \oint_\gamma dy_\mu \,\partial _x^2\delta
^3(x-y)  \nonumber \\
&&+\mathcal{O}\left( \frac 1{m^4}\right) .  \nonumber
\end{eqnarray}
The first term defines what is called the writhing number $w(\gamma )$ of a
curve $\gamma $ \cite{gu,writhe}. It can be connected to the so called
self-linking number $L(\gamma )$ by 
\[
w(\gamma )=L(\gamma )-T(\gamma )\;, 
\]
where $T(\gamma )$ is the twist of the framing bundle used to define $%
L(\gamma )$ \cite{gu,writhe}. In the following, we shall use a technique to
analyse planar curves, in which case the writhing number vanishes\footnote{%
We recall here that $\gamma$ is a smooth closed curve without
self-intersecting points.}.

In order to calculate the mass-dependent corrections, we first establish a
regularization for the delta function through the well-known representation 
\begin{equation}
\delta ^3(x-y)=\lim_{\alpha \rightarrow 0}\frac 1{(2\pi )^3}\;\int
d^3p\;\frac 1{\left( p^2\right) ^\alpha }e^{ip\cdot (x-y)}\;,  \label{delrep}
\end{equation}
and the $\alpha \rightarrow 0$ limit will be taken at the end of the
computation. For the first contribution of order $1/m$ in the eq.(\ref{wlm3}%
) we thus write 
\begin{eqnarray}
\mathcal{J}_\alpha &\equiv &\oint_\gamma dx^\mu \oint_\gamma dy_\mu \;\delta
^3(x-y)\;  \label{ja} \\
&=&\frac 1{(2\pi )^3}\;\int d^3p\;\frac 1{\left( p^2\right) ^\alpha
}\;f_\gamma ^\mu (p)\;f_{\mu \gamma }^{\;*}(p)\;,  \nonumber
\end{eqnarray}
where 
\begin{equation}
\;f_\gamma ^\mu (p)=\oint_\gamma dx^\mu \;e^{ipx}\;,  \label{f}
\end{equation}
is the Fourier transform of the line element. Observe that for closed curves 
\begin{equation}
p_\mu \;f_\gamma ^\mu (p)=0\;.  \label{closed}
\end{equation}
In order to give a more concrete idea of the evaluation of the integral $%
\mathcal{J}_\alpha $ we specify the curve defining the loop. Therefore, we
shall concentrate on the case in which the curve $\gamma $ is a circle of
radius $R$.

Since the curve is planar, we may decompose the momentum variable as in refs.%
\cite{jj,boan}: 
\[
p_\mu =\widehat{p}_\mu +p_\mu ^{\bot }\;, 
\]
where $\widehat{p}_\mu $ is the projection of $p_\mu $ over the plane
containing $\gamma $, and $p_\mu ^{\bot }$ is the orthogonal component to
that plane. From the definition (\ref{f}), we also have that $f_\gamma ^\mu
(p)=f_\gamma ^\mu (\widehat{p})$ and 
\begin{equation}
\varepsilon _{\mu \nu \rho }\;\widehat{p}^\mu f_\gamma ^\nu (\hat{p}%
)\;f_\gamma ^{*\rho }(\widehat{p})=0\;.  \label{planar}
\end{equation}
Thus, 
\[
\mathcal{J}_\alpha =\frac 2{(2\pi )^3}\;\int d^2\widehat{p}\left(
\int_0^\infty dp^{\bot }\;\frac 1{\left( (p^{\bot })^2+\widehat{p}^2\right)
^\alpha }\right) \;\left| f_\gamma (\widehat{p})\right| ^2\;.\; 
\]
The integral in the orthogonal component is evaluated \cite{jj,boan} as 
\[
\int_0^\infty dp^{\bot }\;\frac 1{\left( (p^{\bot })^2+\widehat{p}^2\right)
^\alpha }=\frac 12\left( \hat{p}^2\right) ^{\frac{1-2\alpha }2}\;\frac{%
\Gamma (\alpha -\frac 12)\Gamma (\frac 12)}{\Gamma (\alpha )}; 
\]
also, for a circle of radius $R$ \cite{jj,boan}, 
\[
\left| f_\gamma (\widehat{p})\right| ^2=4\pi ^2R^2J_1^2(\widehat{p}R)\;, 
\]
where $J_1$ is the Bessel function. By performing the angular integration in 
$d^2p$, it follows that 
\begin{equation}
\mathcal{J}_\alpha =\frac{\Gamma (\alpha -\frac 12)\Gamma (\frac 12)}{\Gamma
(\alpha )}\;R^2\int_0^\infty d\widehat{p}\;\widehat{p}^{2-2\alpha }J_1^2(%
\widehat{p}R)\;,  \label{grad}
\end{equation}
where now $\hat{p}$ denotes the radial variable. The solution to the above
integral may be taken from the table \cite{table}, leading to the expression 
\begin{equation}
\mathcal{J}_\alpha =\frac{R^{2\alpha -1}}{2^{2\alpha -2}}\frac{\Gamma
(2\alpha -2)\Gamma (\frac 12)\Gamma (\frac{5-2\alpha }2)}{\Gamma (\alpha
)\Gamma (\frac{2\alpha +1}2)\Gamma (\frac{2\alpha -1}2)}\;.  \label{jota}
\end{equation}
The $\alpha \rightarrow 0$ limit may now be performed, giving finally 
\begin{equation}
\mathcal{J}_0=-\frac 3{4R}\;.  \label{jota0}
\end{equation}
For planar curves, one can show that all even powers in $1/m$ vanish
automatically by making use of eq.(\ref{planar}). Therefore, the next
nonvanishing contribution for $I(\gamma ,\gamma )$ in eq.(\ref{wlm3}) is
that of order $1/m^3$, \textit{i.e.,} 
\begin{equation}
\oint_\gamma dx^\mu \oint_\gamma dy^\nu \;\partial _x^2\delta ^3(x-y)=-%
\mathcal{J}_{\alpha -1}\;,  \label{m3}
\end{equation}
which, using eq.(\ref{jota}), is computed to be 
\[
\mathcal{J}_{\alpha -1}=\frac{R^{2\alpha -3}}{2^{2\alpha -4}}\frac{\Gamma
(2\alpha -4))\Gamma (\frac 12)\Gamma (\frac{7-2\alpha }2)}{\Gamma (\alpha
-1)\Gamma (\frac{2\alpha -1}2)\Gamma (\frac{2\alpha -3}2)}\;. 
\]
Thus, in the limit $\alpha \rightarrow 0$, we obtain 
\begin{equation}
\mathcal{J}_{-1}=\frac{15}{32R^3}\;.  \label{jota1}
\end{equation}
Finally, substituting eqs.(\ref{jota0}) and $($\ref{jota1}) into eq.(\ref
{wlm3}), and remembering that the writhing of a circle is zero, for the
large-mass corrections to $I(\gamma ,\gamma )$ in the case of a circle we
obtain the expansion 
\begin{equation}
I(\gamma ,\gamma )\;=\frac 3{4mR}-\frac{15}{32m^3R^3}+\mathcal{O}(\frac
1{m^5})\;.  \label{m3}
\end{equation}
The higher-order corrections can be evaluated in a similar way and lead to
the general formula 
\begin{equation}
I(\gamma ,\gamma )\;=\sum_{n=0}^{\infty } 
\frac 1{(mR)^{2n+1}}\frac 1{n+1}\frac{\Gamma (\frac{5+2n}2)}{\Gamma (\frac{1-2n}2)}%
\;\;.  \label{allm}
\end{equation}

\section{Conclusion}

We have proven that, in the large-mass limit, the loop factor $\mathcal{W}%
(\gamma ,\gamma ^{\prime })$ evaluated in the Maxwell--Chern--Simons theory
is not affected by $1/m$-corrections, provided the two curves $\gamma
,\gamma ^{\prime }$ do not touch each other.

It is worth underlining that this result has been achieved by means of the
field redefinition (\ref{gen-red}), which turns out to provide a very useful
computational tool in order to deal with the large-mass dependence of loop
variables in three-dimensional gauge theories, including the case of the
Wilson loop.

\appendix 

\section{Small Mass Expansion}

In this appendix we discuss briefly, along the lines developed in the
present article, the complementary question of the small-mass behaviour of
the link variable 
\begin{equation}
I(\gamma ,\gamma ^{\prime })=\oint_\gamma dx^\mu \oint_{\gamma ^{\prime
}}dy^\nu \left\langle A_\mu (x)A_\nu (y)\right\rangle _{\text{MCS}}
\label{link}
\end{equation}
in the context of the Maxwell--Chern--Simons theory.

For this purpose, we make use of another kind of field redefinition which
now allows to reabsorb the Chern--Simons term into the Maxwell one in the
action, that is, 
\begin{equation}
\frac 12\int d^3x\,\varepsilon ^{\mu \nu \rho }A_\mu \partial _\nu A_\rho
+\frac 1{4m}\int d^3x\,F_{\mu \nu }F^{\mu \nu }\;=\frac 1{4m}\int d^3x\,%
\widehat{F}_{\mu \nu }\widehat{F}^{\mu \nu }\;,  \label{ym-red}
\end{equation}
through a suitable gauge-field transformation of the type 
\begin{equation}
A_\mu =\widehat{A}_\mu +\sum_{n=1}^\infty m^n\widehat{\theta }_\mu ^n\;,
\label{yma-red}
\end{equation}
in which the first few coefficients $\hat{\vartheta}_\mu ^n$ are computed to
be 
\begin{eqnarray}
\widehat{\vartheta }_\mu ^1 &=&\frac 14\varepsilon _{\mu \nu \rho }\frac
1{\partial ^2}F^{\nu \rho }\;,\;\;\;\;\;\;\;\;\;\;\;\;\;\;\;\;\;\;\widehat{%
\vartheta }_\mu ^2=-\frac 38\frac 1{\partial ^4}\partial ^\nu F_{\mu \nu }\;,
\label{four} \\
\;\widehat{\vartheta }_\mu ^3 &=&\frac 5{32}\varepsilon _{\mu \nu \rho
}\frac 1{\partial ^4}F^{\nu \rho }\;,\;\;\;\;\;\;\;\;\;\;\;\;\;\;\;\;%
\widehat{\vartheta }_\mu ^4=\frac{13}{64}\frac 1{\partial ^6}\partial ^\nu
F_{\mu \nu }\;.  \nonumber
\end{eqnarray}
We observe that, like in the redefinition of Sect.2, the $\hat{\vartheta}%
_\mu ^n$'s are gauge invariant and transverse.

However they are nonlocal, as may be inferred from the presence of the
inverse of the laplacian. This feature will spoil the integral $I(\gamma
,\gamma ^{\prime })$ in eq.(\ref{link}) of any topological meaning. As one
can easily understand, this is due to the fact that the small-mass behaviour
is dominated by the pure Maxwell term which, of course, is not of a
topological nature.

Such a change of variables leads to a computation of the link variable
within the pure Maxwell theory: 
\[
I(\gamma ,\gamma ^{\prime })=\oint_\gamma dx^\mu \oint_{\gamma ^{\prime
}}dy^\nu \left\langle A_\mu (\widehat{A}(x))A_\nu (\widehat{A}%
(y))\right\rangle _{\text{Maxwell}}. 
\]
We obtain, therefore, a small-mass expansion for $I(\gamma ,\gamma ^{\prime
})$, whose first constributions are 
\begin{eqnarray}
I(\gamma ,\gamma ^{\prime }) &=&\oint_\gamma dx^\mu \oint_{\gamma ^{\prime
}}dy^\nu \left\langle \widehat{A}_\mu (x)\widehat{A}_\nu (y)\right\rangle _{%
\text{Maxwell}}  \nonumber \\
&&-\frac m{4\pi }\int d^3z\oint_\gamma dx^\mu \oint_{\gamma ^{\prime
}}dy^\nu \,\,\varepsilon _{\nu \alpha \beta }\frac 1{\left| y-z\right|
}\partial _z^\alpha \left\langle \widehat{A}_\mu (x)\widehat{A}^\beta
(z)\right\rangle _{\text{Maxwell}}  \nonumber \\
&&+\mathcal{O}(m^2).  \label{final}
\end{eqnarray}
We may now substitute in the above expression for the Maxwell propagator, 
\begin{equation}
\left\langle \widehat{A}_\mu (x)\widehat{A}_\nu (y)\right\rangle _{\text{%
Maxwell}}=\frac m{4\pi }\frac 1{\left| x-y\right| }g_{\mu \nu },
\label{ym-prop}
\end{equation}
with the result 
\begin{eqnarray}
I(\gamma ,\gamma ^{\prime })\; &=&\frac m{4\pi }\oint_\gamma dx^\mu
\oint_{\gamma ^{\prime }}dy^\nu \;\frac 1{\left| x-y\right| }g_{\mu \nu } 
\nonumber  \label{final1} \\
&&+\left( \frac m{4\pi }\right) ^2\int d^3z\oint_\gamma dx^\mu \oint_{\gamma
^{\prime }}dy^\nu \,\varepsilon _{\mu \nu \alpha }\frac 1{\left| y-z\right|
}\partial _x^\alpha \frac 1{\left| x-z\right| }  \label{final1} \\
&&+\mathcal{O}(m^3).  \nonumber
\end{eqnarray}

Two remarks are in order. First, we see that, as already underlined, the
topological character of $I(\gamma ,\gamma ^{\prime })$ is lost in the
small-mass regime. Second, the contributions may be evaluated only after
specifying the curves $\gamma $ and $\gamma ^{\prime }$.

Let us conclude by underlining that eq.(\ref{final1}) also applies to the
case of the Wilson loop (\ref{wilsonl}). For instance, for the first
contribution of the expansion (\ref{final1}), we get 
\begin{equation}
I(\gamma ,\gamma )\;=\frac m{4\pi }\oint_\gamma dx^\mu \oint_\gamma dy_\mu
\;\frac 1{\left| x-y\right| }\;+\;\mathcal{O}(m^2)\;.  \label{smexp}
\end{equation}
In the case of the circle, the above integral can be evaluated by following
the same procedure of Sect.5, yielding 
\begin{eqnarray}
I(\gamma ,\gamma ) &=&m\mathcal{J}_{\alpha +1}\;+\;\mathcal{O}(m^2)
\label{smexp-res} \\
&=&m\frac{R^{2\alpha +1}}{2^{2\alpha }}\frac{\Gamma (2\alpha )\Gamma (\frac
12)\Gamma (\frac{3-2\alpha }2)}{\Gamma (1+\alpha )\Gamma (\frac{2\alpha +3}%
2)\Gamma (\frac{2\alpha +1}2)}\;+\;\mathcal{O}(m^2)\;.  \nonumber
\end{eqnarray}
As expected, the limit $\alpha \rightarrow 0$ is singular, due to the
presence of $\Gamma (2\alpha )$. We have recovered thus the well-known
divergent contribution to the Wilson loop of the pure Maxwell term in three
dimensions \cite{jj,boan,rr}.

\vspace{2cm}

{\Large \textbf{Acknowledgments}}

S.P. Sorella wishes to dedicate this work to the memory of his friend and
colleague Liviu Tataru.

We are grateful to D.G. Barci for useful discussions and for pointing out to
us ref.\cite{br}. The Conselho\ Nacional de Pesquisa e Desenvolvimento (CNPq
/Brazil), the Funda\c {c}\~{a}o de Amparo \`{a} Pesquisa do Estado do Rio de
Janeiro (Faperj) and the SR2--UERJ are gratefully acknowledged for financial
support.

\end{document}